\documentclass[pdftex,modern]{aastex63}
\usepackage{multirow}
\usepackage{color}

\newcommand {\sm}{\mbox{$M_{\odot}$}}
\newcommand {\kms}{\mbox{km~s$^{-1}$}}
\newcommand {\cooz}{\mbox{$^{12}$CO($J$~=~1--0)}}
\newcommand {\coto}{\mbox{$^{12}$CO($J$~=~2--1)}}
\newcommand {\cott}{\mbox{$^{12}$CO($J$~=~3--2)}}

\newcommand {\vl}{\mbox{$V_\mathrm{lsr}$}}

\newcommand {\nhtwo}{\mbox{$N_\mathrm{H_2}$}}

\newcommand{\ratio}{\mbox{$R_{3-2/1-0}$}}

\received{January 31, 2022}
\accepted{Febrary 25, 2022}
\submitjournal{ApJ}

\shorttitle{ALMA observations of the central region of NGC~253}
\shortauthors{Konishi et al.}

\begin{document}
\title{Discovery of a giant molecular loop in the central region of NGC~253}

\correspondingauthor{R. Konishi}
\email{s\_r.konishi@p.s.osakafu-u.ac.jp}

\author[0000-0003-0763-1529]{R. Konishi}
\affiliation{Department of Physical Science, Graduate School of Science, Osaka Prefecture University, 1-1 Gakuen-cho, Naka-ku, Sakai, Osaka 599-8531, Japan}

\author[0000-0003-2735-3239]{R. Enokiya}
\affiliation{Department of Physics, Institute of Science and Technology, Keio University, 3-14-1 Hiyoshi, Kohoku-ku, Yokohama, Kanagawa 223-8522, Japan}

\author[0000-0002-8966-9856]{Y. Fukui}
\affiliation{Department of Physics, Nagoya University, Furo-cho, Chikusa-ku, Nagoya, Aichi, 464-8601, Japan}

\author[0000-0002-3373-6538]{K. Muraoka}
\affiliation{Department of Physical Science, Graduate School of Science, Osaka Prefecture University, 1-1 Gakuen-cho, Naka-ku, Sakai, Osaka 599-8531, Japan}

\author[0000-0001-8969-4237]{K. Tokuda}
\affiliation{National Astronomical Observatory of Japan, Mitaka, Tokyo 181-8588, Japan}
\affiliation{Department of Physical Science, Graduate School of Science, Osaka Prefecture University, 1-1 Gakuen-cho, Naka-ku, Sakai, Osaka 599-8531, Japan}

\author[0000-0001-7826-3837]{T. Onishi}
\affiliation{Department of Physical Science, Graduate School of Science, Osaka Prefecture University, 1-1 Gakuen-cho, Naka-ku, Sakai, Osaka 599-8531, Japan}

\begin{abstract}
NGC~253 is a starburst galaxy of SAB(s)c type with increasing interest because of its high activity at unrivaled closeness.
Its energetic event is manifested as the vertical gas features in its central molecular zone, for which stellar feedback was proposed as the driving engine.
In order to pursue details of the activity, we have undertaken a kinematic analysis of the ALMA archive data of $\cott$ emission at the highest resolution $\sim$3 pc.
We revealed that one of the non-rotating gas components in the central molecular zone shows a loop-like structure of $\sim$200 pc radius.
The loop-like structure is associated with a star cluster, whereas the cluster is not inside the loop-like structure and is not likely as the driver of the loop-like structure formation.
Further, we find that the bar potential of NGC~253 seems to be too weak to drive the gas motion by the eccentric orbit.
As an alternative, we frame a scenario that magnetic acceleration by the Parker instability is responsible for the creation of the loop-like structure.
We show that the observed loop-like structure properties are similar to those in the Milky Way, and argue that recent magneto-hydrodynamics simulations lend support for the picture having the magnetic field strength of $\gtrsim$100 $\mu$G.
We suggest that cluster formation was triggered by the falling gas to the footpoint of the loop, which is consistent with a typical dynamical timescale of the loop $\sim$1 Myr. 

\end{abstract}


\section{Introduction} \label{sec:intro}

Star formation activity heats up and ionize the surrounding gas and promote galactic evolution thereby.
Because star formation takes place in the molecular clouds, observations of the molecular gas is a crucial step in understanding galactic evolution.
In the Milky Way (MW), the gravitational potential of the galactic center within 1 kpc is about 10 times deeper than in the galactic disk, and the material in the galactic center is subject to stronger compression.
It is possible that such compression affects star formation in the galactic center but not so in the galactic disk.
Therefore, it is crucial to understand the physical conditions like gas density, gas temperature, magnetic field, etc. in order to elucidate galactic evolution.
It is also known that the gas temperature and the gas velocity dispersion are 10 times higher in the galactic center than in the galactic disk \citep[e.g.,][]{mor96, mar04}, whereas their origin remains veiled.

The starburst galaxies show extremely active star formation and the feedback effects including the stellar winds and supernova explosions (SNe) influence the surroundings dynamically \citep[e.g.,][]{sak06, bol13}.
NGC~253, in particular, is one of the nearby galaxies is forming a number of high-mass stars in the central region \citep[e.g.,][]{kkor09, ben15, ler15, ler18, dav16, and17, man19, coh20, ric20, lev21, mil21} and discoveries of loop-like structures and supershells that are formed by stellar feedback \citep[e.g.,][]{sof94, sak06} and starburst outflow \citep[][]{bol13, wal17, zsc18, kri19, lev21, mil21} have been reported.

In addition, several dense clumps, which dominate the starburst activity in the central region are embedded in an extended, non-axisymmetric, bar-like gas distribution in the central region of NGC~253 disk \citep[][]{ler15}.
\citet[][]{kri19} suggested that the central region of NGC~253 disk consisted of the main disk ("disk") and the additional kinetic disk/bar component which is not consistent with rotation of the main disk ("otherdisk").

Meanwhile \citet[][]{sak11} suggested that the spatial distribution of the Central Molecular Zone (CMZ) of NGC~253 is similar to the CMZ of the MW.
\citet[][]{sak11} found that the CMZ of NGC~253 has the FWHM spatial size of 300 pc $\times$ 100 pc and the several prominent CO peaks of the size of 20--40 pc based on $\coto$ observation at $\sim$1$\arcsec$.
The authors suggest that the spatial size is similar to that of the CMZ of the MW and the CO peaks are corresponding to Sgr~A, Sgr~B2, Sgr~C, and $l$=1$^{\circ}$.3 complex.

The star forming activities in the CMZs of the MW and NGC~253 largely differ from each other.
The star formation rate in the CMZ of the MW is estimated to be $\sim$0.02--0.1 $\sm$ yr$^{-1}$ \citep[][]{yus09, imm12, lon13}, whereas that in the CMZ of NGC~253 is one to two orders of magnitude higher \citep[$\sim$ 1.7--2.8 $\sm$ yr$^{-1}$;][]{ott05, ben15, ler15}.
The large difference of the power of dynamical engines at the both galaxies' hearts is expected to invoke significantly different gas dynamics.
However, gas distributions of the CMZs of two galaxies are very similar according to \citet{sak11}.
In order to explore this apparent inconsistency, detailed analyses of gas dynamics in the CMZ of NGC~253 are most important.

Similar non-rotating components are observed in the CMZ of the MW and are interpreted as due to the bar potential \citep[e.g.,][]{bin91} or the floatation by the magnetic instabilities \citep[e.g.,][]{fuk06}.

In the central region of the MW, there are two families of orbits called $x_{1}$ and $x_{2}$ in the reference frame of the small bar. 
The $x_{1}$ orbits are elongated along the major axis of the small bar, whereas $x_{2}$ orbits are distributed within the innermost $x_{1}$ orbits and elongated along the minor axis of the small bar.
The gas velocities evoked by these two orbits produce non-circular motions.

Recent numerical simulations reproduce the parallelogram in the $l$--$v$ diagram, which is consistent with the observations \citep[e.g,][]{bin91, rod06}.
So far, there is some indirect evidence for the small bar based on far-infrared observations \citep[e.g.,][]{mol11}, but a direct evidence has not been reported yet.

Magnetic instabilities are another candidate for the non-circular acceleration.
In regions with lower gas density caused by the fluctuation of gas density, the magnetic loop can float because the buoyancy at the top of the loop exceeds the magnetic tension by Parker instability \citep[][]{par66}.
The gas on the loop bridge slides down along the magnetic loop, and produces velocity gradient.
Since such gas collides supersonically at the loop footpoints, the velocity dispersion at the footpoints is enhanced \citep[][]{eno21}.

Several studies so far have been reported in total five magnetic flotation loops in the center of the MW \citep[][]{fuk06, fuj09, tor10, tor10_2, kud11, eno21}, and the authors suggest that the loops contribute to high-velocity dispersion in the center of the MW.
Some studies based on magneto-hydrodymamics simulations show that magnetic loops are reproduced in the central region of galaxies \citep[][]{mat88, mac09, suz15}.

This situation is being overcome thanks to ALMA, which allows us to resolve molecular clouds at 5 pc resolution on NGC~253.
This paper presents an analysis of the gas kinematics in the central region of NGC~253 using the ALMA Cycle 1 $\cooz$ and Cycle-3 $\cott$ data, and explores the origin of the gas with non-circular motions in the region.
$\cooz$ was used to derive the total gas mass and the higher excitation transition line $\cott$ was used to reveal detailed gas kinematics.
We adopt a distance of NGC~253 to be $\sim$3.5 Mpc, where 1$\arcsec$ corresponds to $\sim$17 pc \citep[][]{rek05} and the galactic center was taken as
J2000 0h47m33.14s, $-$25d17m17.52s \citep[][]{mul10} and its velocity $\sim$243 $\kms$ \citep[][]{kor04}.
We give the observational data in Section~\ref{sec:obs} and the results in Section~\ref{sec:results}, discussion on the gas with non-circular motion is made in Section~\ref{sec:dis}, and conclusions are given in Section~\ref{sec:conclusion}.

\section{Observations and data reduction} \label{sec:obs}

\subsection{Data sets}\label{obs:space}

The data sets were observed toward the CMZ of NGC~253 with ALMA under Cycle 1 in Band 3 (project code: 2012.1.00108.S, PI: A. Bolatto) and under Cycle 3 in Band 7 (project code: 2015.1.00274.S, PI: A. Bolatto). 
The details of the data sets are presented in Table~\ref{tab:table1}.

In the data set in Band 3, the lower side band (LSB) is tuned to 101.7--105.4 GHz and the upper side band (USB) to 118.8--113.7 GHz and 113.8--115.6 GHz, which covers the frequency of $\cooz$ emission. 
The total bandwidth is 7.5 GHz and the frequency resolution is 488.3 kHz (corresponding to the velocity resolution of $\sim$3 $\kms$). 
The data set consists of the 12 m array (C32-2) with seven pointings at $\sim$2 kpc of the CMZ of NGC~253, the ACA with three pointings at $\sim$2 kpc of the CMZ of NGC~253, and the total power (TP) array. 
The baseline of the 12 m array and ACA are 15--463 m and 7--49 m respectively.
The calibrator were J0334-4008 for bandpass, J0038-2459 for phase and Uranus for flux. 
The visibility of the 12 m configuration data is calibrated using Common Astronomy Software Application (CASA) 4.2.2 and the calibration script prepared by the ALMA observatory. 
The visibility of the ACA is calibrated using CASA 4.2.0 and the calibration script prepared by the ALMA observatory.

In the data set in Band 7, the LSB to 342.0--345.8 GHz is tuned, which includes $\cott$ emission line, and the USB to 353.9--357.7 GHz. 
The total bandwidth and the frequency resolution is 7.5 GHz and 976.6 kHz, respectively (corresponding to the velocity resolution of $\sim$0.8 $\kms$). 
The data set consists of a 12 m extended configuration and the 12 m compact configuration (hereafter 12E and 12C) with four pointings at $\sim$0.8 kpc toward the center of NGC~253, the ACA with five pointings at $\sim$1.0 kpc of the center of NGC~253 and TP. 
The baseline of the 12E, 12C and the ACA are 15--1813 m, 15--784 m, and 9--49 m, respectively.
The calibrator were J0006-0623 for bandpass, J0038-2459 for phase, and Pallas for flux.
The visibility of 12C is calibrated using CASA 4.6.0 pipeline and the calibration script prepared by the ALMA observatory.
The visibility of the others are calibrated using CASA 4.7.2 pipeline and the calibration script prepared by the ALMA observatory.

\subsection{Imaging}\label{obs:image}

For $\cooz$, we subtracted the continuum component from the 12 m configuration and the ACA visibilities using the task of \texttt{uvcontsub} in CASA 5.4.0 and we ran the task of \texttt{tclean} to image cube map of the 12 m configuration and the ACA with the Briggs weighting scheme with a \texttt{robust} parameter of 0.5. 
We combined the 12 m configuration, the ACA and TP by using the task of \texttt{feather}, and we obtained a $\cooz$ cube with an angular resolution of 1$\farcs$6 $\times$ 1$\farcs$2. 
The data was gridded to be the spatial grid size of 0$\farcs$3 and a velocity grid size of 5 $\kms$.

For $\cott$, we did not use the task of \texttt{uvcontsub} for subtracting continuum component because there exist a lot of  molecular lines within the frequency range of $\cott$ (e.g., HC$_{3}$N, NS, CH$_{3}$OH, SO$_{2}$).
Alternatively, we ran the task of \texttt{tclean} to make a cube of 12E, 12C, and ACA with the Briggs weighting scheme with \texttt{robust} parameter of 0.5, and we subtracted the continuum component using the task of \texttt{imcontsub}.
We combined all four data sets by using the task of \texttt{feather}, and finally obtained a $\cooz$ cube with the angular resolution of 0$\farcs$18 $\times$ 0$\farcs$14. 
The data was gridded to be the spatial size of 0$\farcs$03 and the velocity element of 2 $\kms$ .

\begin{deluxetable*}{ccc}
\tabletypesize{\scriptsize}
\tablewidth{0pt} 
\tablecaption{Details of the datasets
\label{tab:table1}}
\tablehead{
\colhead{}                  & \colhead{$\cooz$}  & \colhead{$\cott$}  
\\[-2.5mm]
}
\startdata 
 ALMA project code & 2012.1.00108.S & 2015.1.00274.S \\
 Configurations & ALMA 12 m array and ACA & \multicolumn{1}{p{4cm}}{ALMA 12 m extend array, 12 m compact array and ACA} \\
 TP & ALMA TP & ALMA TP \\
 Angular resolution & $\sim$1$\farcs$6 $\times$ 1$\farcs$2 & $\sim$0$\farcs$18 $\times$ 0$\farcs$14 \\
Velocity resolution & 5 $\kms$ & 2 $\kms$ \\
Line sensitivity & $\sim$0.08 K & $\sim$0.4 K \\
\enddata
\tablecomments{TP is a total power array that covers the largest spatial distribution of emission lines. We inferred an angular and velocity resolution from the images, which are not smoothed. We derived line sensitivity from the root mean square of noise at non-emission regions.}
\end{deluxetable*}

\subsection{Rotating position angle of the CMZ of NGC~253}\label{obs:rot}

In this paper, we focus on gas dynamics in the CMZ of NGC~253.
In order to compare height of the gas in the CMZ of NGC~253, we adjusted the position angle and velocity of the data.
We rotated the position angle of 55$^{\circ}$ \citep[][]{ler15, kri19} to 90$^{\circ}$.
Figure~\ref{fig:fig_chn_wholedisk} shows gas distributions in equatorial coordinates to compare those identified by the previous studies.
The subsequent figures show the processed data.

\section{Results}\label{sec:results}

\subsection{Distribution of high excitation gas}\label{result:distribution}
We investigate the spatial-velocity distribution of molecular gas with the higher resolution and higher transition gas tracer, $\cott$.

\begin{figure}[htbp]	
\includegraphics[width=155mm]{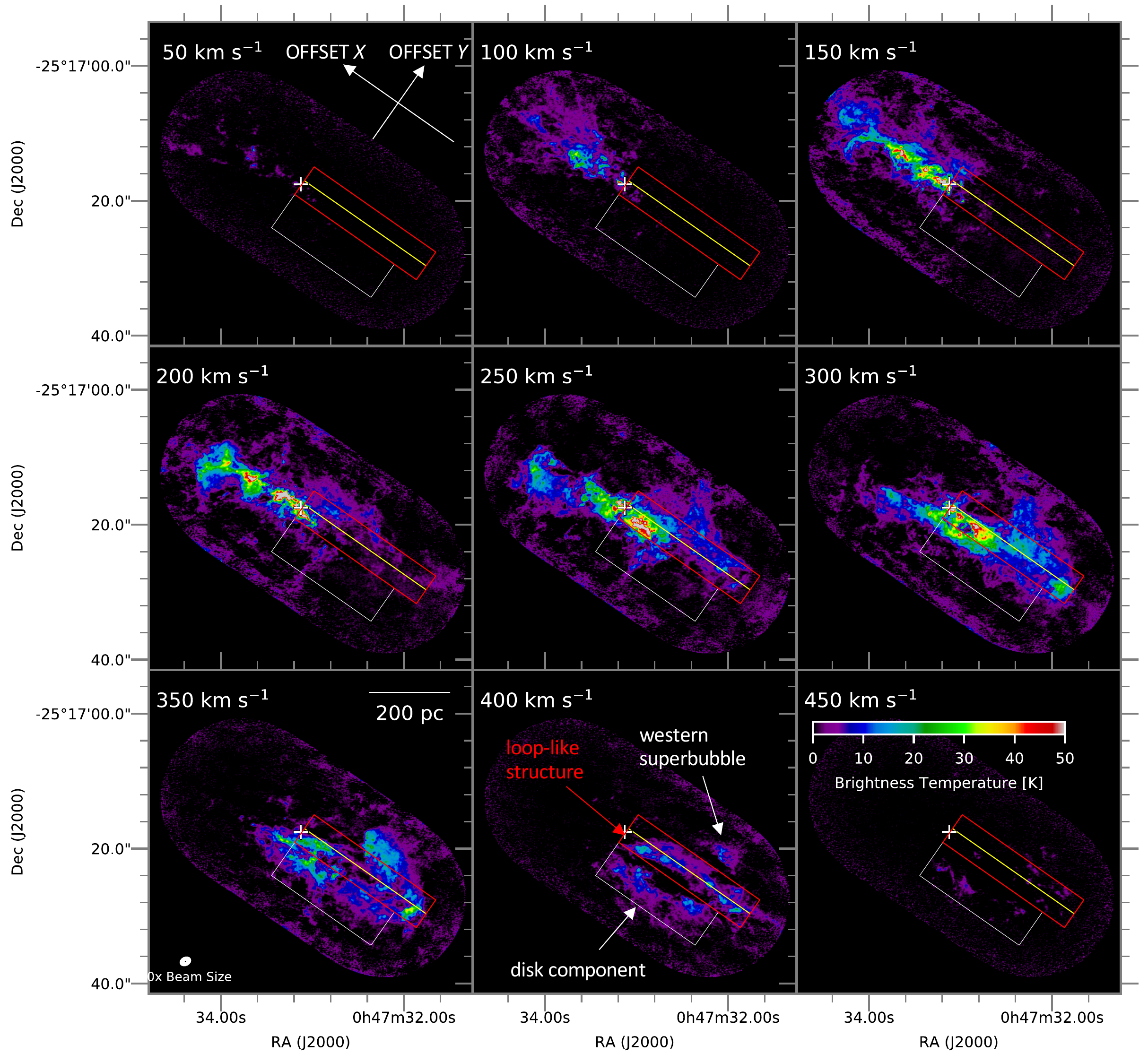}
\caption{Velocity channel map of $\cott$ toward the CMZ of NGC~253.
The white cross indicates the position of the center of NGC~253 \citep[][]{mul10}.
Directions of OFFSET $X$ and OFFSET $Y$ axes are shown at upper right corner of 50 $\kms$ channel, and the intersection of these axes is corresponding to the center of NGC~253.
The original synthesized beam (black) and the 10 times zoomed beam (white) are indicated at the lower left corner of the 350 $\kms$ channel.
The region of the loop-like structure and the disk component at 400 $\kms$ are indicated by the red and white rectangles, respectively.
A yellow line shows an intermediate line of the short side of the red rectangle.
The position of the western superbubble \citep[][]{kri19} is shown in the panel at $\vl$ = 400 $\kms$.
\label{fig:fig_chn_wholedisk}}
\end{figure}

Figure~\ref{fig:fig_chn_wholedisk} shows velocity channel distribution of $\cott$ emission from $\vl$ = 100 to 450 $\kms$.
A main body of molecular gas is seen in the northeast of the galactic center at $\vl$ = 100 $\kms$, and this gas extends over the galactic center at $\vl$ = 150 $\kms$ to 200 $\kms$.
The gas centered on the galactic center and symmetrically spreads over more than 300 pc from the northeast to southwest at the vicinity of systemic velocity of the galaxy, 243 $\kms$.
The gas moves to the southwest in $\vl$ = 300--400 $\kms$ and disappears toward the southwest at $\vl$ = 450 $\kms$.
As a whole, the gas with brightness temperature above 10 K exhibits a clear rotating disk, linearly extends from the northeast to southwest of the galactic center, and blueshifts in the northeast and red-shifts in the southwest. 
We hereafter call this the main gas stream disk component.

At $\vl$ = 400 $\kms$, three molecular structures are seen.
The northern structure is a part of the western superbubble discovered by several studies \citep[][]{sak06, bol13, kri19} and the southern one is a part of the disk component that shows the galactic rotation and connects to the galactic center gas at $\vl$ = 300--350 $\kms$.
The central gas structure indicated by the red rectangle in Figure~\ref{fig:fig_chn_wholedisk} appears at $\vl$ = 350 $\kms$ and shows a loop-like structure, which extends from the vicinity of the galactic center and connects to the disk component at its southwest end at $\vl$ = 400 $\kms$.
The disk component is located toward the same region at $\vl$ = 250--300 $\kms$, meaning the loop-like structure has higher velocity than the disk.
The length, width, and height of the loop-like structure are $\sim$300 pc, $\sim$50 pc, and $\sim$20 pc, respectively.
For the following, we focus on this loop-like structure showing a peculiar velocity.

\begin{figure}[htbp]	
\centering
\includegraphics[width=150mm]{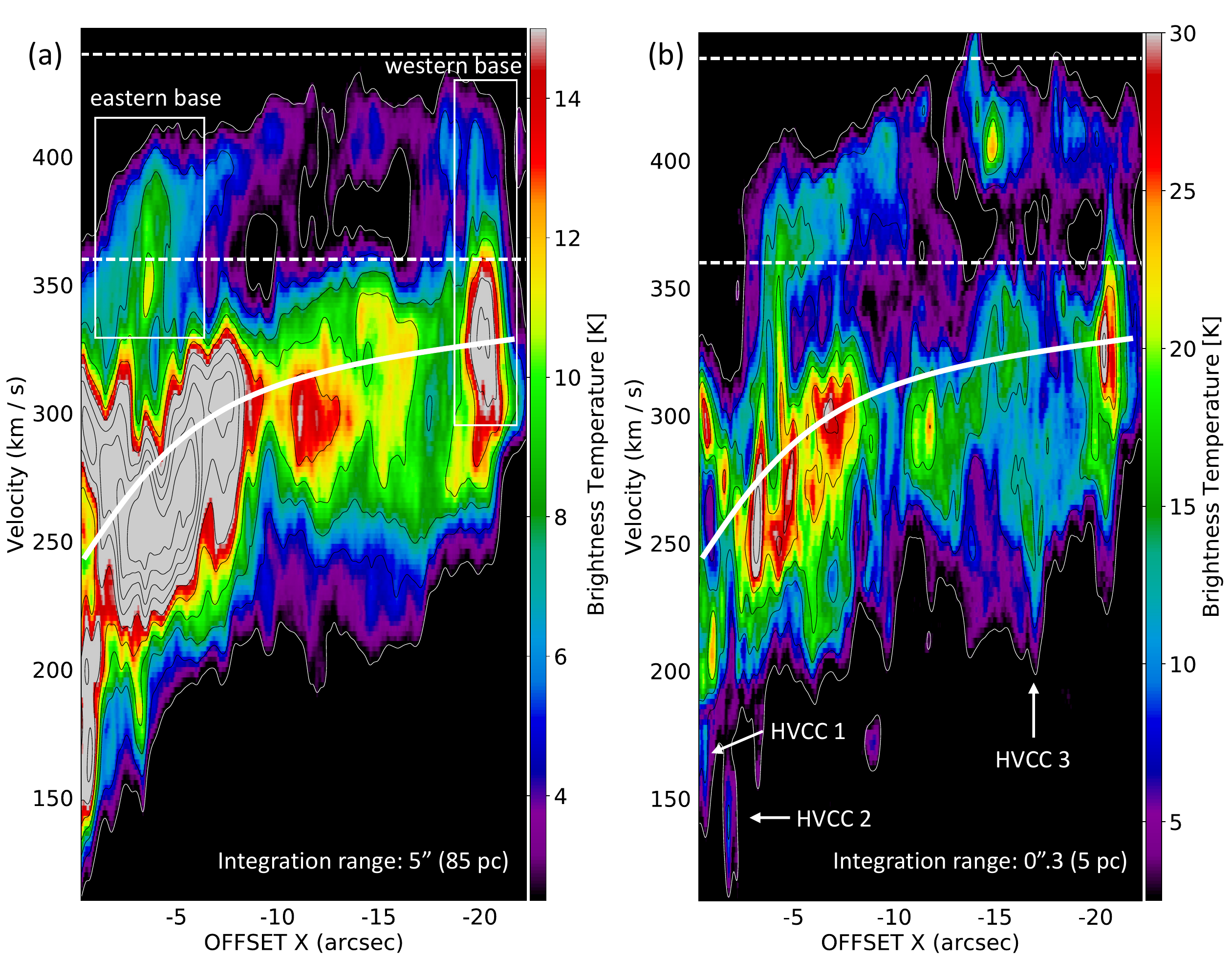}
\caption{(a) Position-velocity diagram of $\cott$ integrated whole the loop-like structure width of 85 pc.
The axis of OFFSET $X$ is horizontal the position angle of the CMZ of NGC~253 of $\sim$55$^{\circ}$\citep[][]{ler15, kri19}. A white solid line indicates the circular rotational velocity component delineated by eye fit.
Dashed lines indicate velocity range for the loop-like structure.
Two rectangles indicate eastern and western bases.
Contours are drawn at 2.5, 5.0, 7.5, 10.0, 12.5, 15.0, 17.5, 20.0, 22.5, and 25.0 K. (b) Position-velocity diagram of $\cott$ toward the direction of the loop-like structure's center with the integrated slice width of 5 pc.
The white solid line and dashed lines are same as Figure~\ref{fig:fig_CO32_pv}a. HVCCs 1--3 indicate the directions of high-velocity cloud components.
\label{fig:fig_CO32_pv}}
\end{figure}

Figures~\ref{fig:fig_CO32_pv}a and \ref{fig:fig_CO32_pv}b show position-velocity (p-v) diagrams of $\cott$ emission toward the loop-like structure with the integrated ranges corresponding to the red rectangle (the whole the loop-like structure) and the yellow line (a slice of the center of the loop-like structure) in Figure~\ref{fig:fig_chn_wholedisk}, respectively.
The OFFSET $X$ axis defined as the offset in the direction of the disk elongation from the galactic center and has the length of $\sim$370 pc.
We selected an emission-free area the same size of the red rectangle and made a calculation of the RMS noise level (1$\sigma \sim$1 K for its position-velocity diagram.
The lowest contour in Figure~\ref{fig:fig_CO32_pv}a corresponds 2.5 K.
In Figure~\ref{fig:fig_CO32_pv}(a) the rotating disk component is seen as a main global structure in $\vl$ $\sim$100--360 $\kms$.
On the other hand, the loop-like structure shows higher velocity of $\vl$ $\sim$360--440 $\kms$ and is clearly located out of range of the disk component.
The loop-like structure shows large velocity width toward both its bases (OFFSET $X$ $\sim$$-$20$\arcsec$ and $\sim$$-$3$\arcsec$) and exhibits a linear velocity increase from OFFSET $X$ = $-$18$\arcsec$ to $-$3$\arcsec$.
We defined the velocity widths of the bases from the velocity range of the clumps corresponding to the bases in Figure~\ref{fig:fig_CO32_pv}a.
We also calculated the RMS noise level for Figure~\ref{fig:fig_CO32_pv}b and derived it to be $\sim$1 K.
The lowest contour in Figure~\ref{fig:fig_CO32_pv}b corresponds 2.5$\sigma$ ($\sim$2.5 K).
We again confirmed the same velocity trend in the loop-like structure, the large velocity width toward the both bases and linear velocity gradient along the middle region, in the narrow integration range, meaning the trend purely comes from the loop-like structure itself rather than an accidental superimpositions of other gas in the same line of sight.

We find a few compact clouds with a large velocity width ($\sim$50 $\kms$) toward OFFSET $X$ = $-$17$\arcsec$, $-$1$\farcs$8, $-$0$\farcs$63 indicated by the white arrows in Figure~\ref{fig:fig_CO32_pv}b.
This may suggest that lots of compact, $\sim$5--10 pc, and large velocity width clouds are embedded in the CMZ of NGC~253.
Similar clouds are identified as High Velocity Compact Clouds \citep[HVCCs;][]{oka98} in the CMZ of the MW, and this may suggest that similar gas dynamics between the CMZ of NGC~253 and the MW CMZ.
\citet[][]{sak11} suggested that the CMZs of the MW and NGC~253 resemble in terms of the FWHM spatial size of 300 pc $\times$ 100 pc in $\coto$ and the several prominent CO peaks of the size of 20 – 50 pc which are corresponding to Sgr~A, Sgr~B2, Sgr~C, and $l$=1$^{\circ}$.3 complex.
It is interesting that even though the star formation activities in the CMZs are quite different between two galaxies, similar HVCCs are discovered in both CMZs.
We summarized positions, FWHM sizes, velocity ranges, molecular masses, and kinetic energy of the HVCCs in Table~\ref{tab:hvccs}.

In order to clearly illustrate the peculiarity of the loop-like structures' position and velocity compared to the global rotating disk component, we corrected the position angle to 90$^{\circ}$ from the data sets.
The new coordinates are described as offsets from the galactic center, namely OFFSET $X$ and OFFSET $Y$.

\begin{figure}[htbp]	
\centering
\includegraphics[width=100mm]{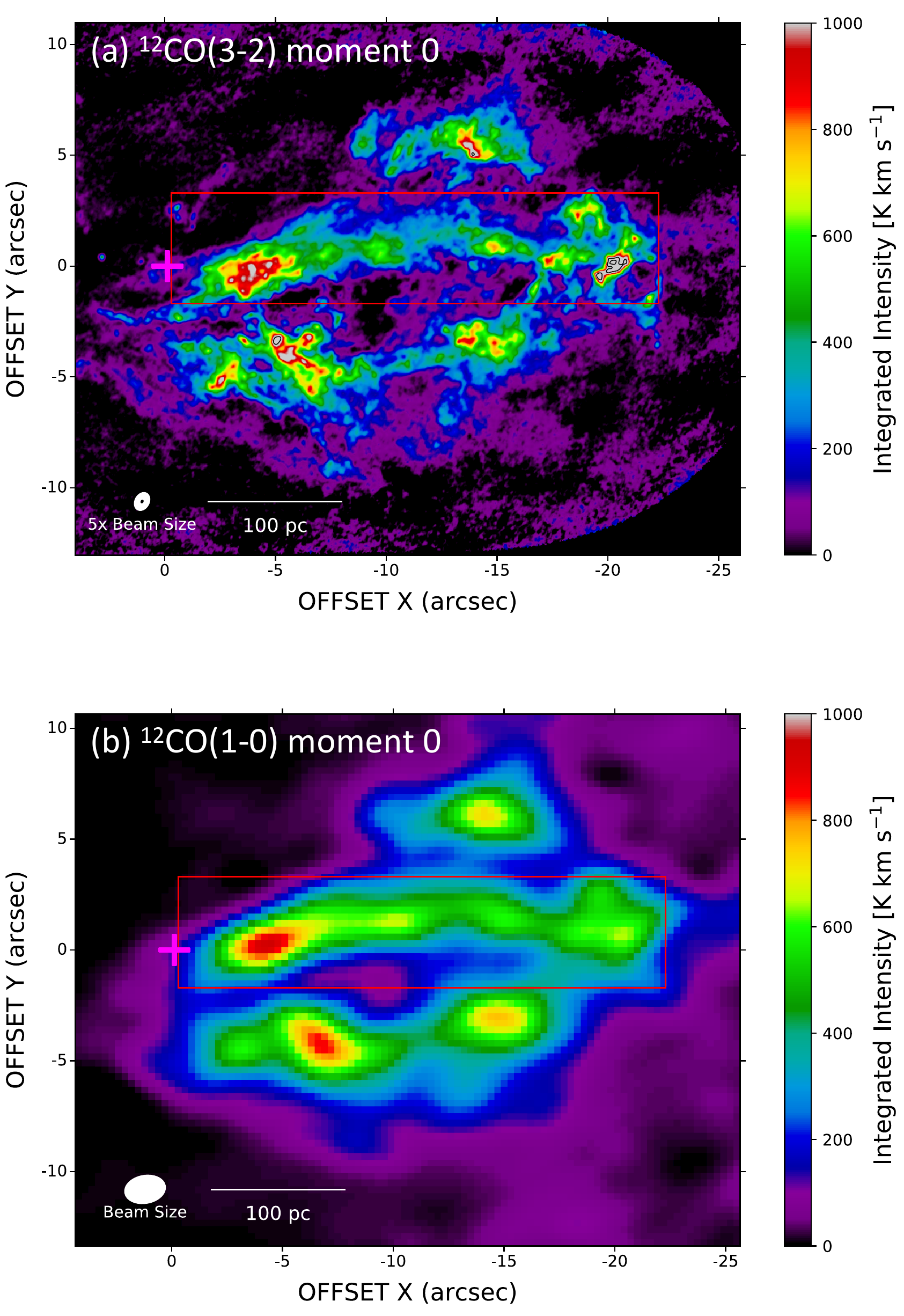}
\caption{(a) Integrated intensity map of $\cott$ integrated from $\vl$ = 360--440 $\kms$.
A pink cross indicates the position of the center of NGC~253.
Contours are drawn at 1200 and 1600 K.
A red rectangle is same as Figure~\ref{fig:fig_chn_wholedisk}.
(b) Same as (a) but for the $\cooz$ map.
\label{fig:fig_loop_rot35}}
\end{figure}

Figures~\ref{fig:fig_loop_rot35}a and \ref{fig:fig_loop_rot35}b show integrated intensity distributions of the loop-like structure integrating over $\vl$ = 360--440 $\kms$ in $\cott$ and $\cooz$, respectively.
The loop-like structure is located $-$22$\farcs$3 $\lesssim$ OFFSET $X$ $\lesssim$ $-$0$\farcs$3, $-$1$\farcs$7 $\lesssim$ OFFSET $Y$ $\lesssim$ 3$\farcs$3.
Figures~\ref{fig:fig_loop_rot35}a and \ref{fig:fig_loop_rot35}b show the loop-like structure has higher integrated intensities at the both ends, and the middle part, which connects two bases, bends toward the OFFSET $Y$ direction.

Although the eastern base of the loop-like structure is located 50 pc away from the galactic center in projection and the velocity of the loop-like structure is far from the disk component, the loop-like structure physically connects to the main gas stream disk component at the eastern base (see Figure~\ref{fig:fig_CO32_pv}a), and hence the loop-like structure is physically connected to the galactic center.
The noise levels for $\cott$ and $\cooz$ intensities are $\sigma \sim$13 K $\kms$ and 7 K $\kms$, respectively.

Figure~\ref{fig:fig_loop_rot35}a shows that the highest intensities above 1000 K $\kms$ are seen at both bases of the loop-like structure.
The FWHM sizes of the bases are $\sim$50 pc and $\sim$20 pc, respectively, from east to west.
Despite an order-of-magnitude coarser angular resolution, $\cooz$ emission also shows loop-like structure (Figure~\ref{fig:fig_loop_rot35}b)
The lower integrated intensity toward the western base in $\cooz$ are possibly originated from the beam dilution due to the poor angular resolution.
\begin{figure}[htbp]	
\centering
\includegraphics[width=150mm]{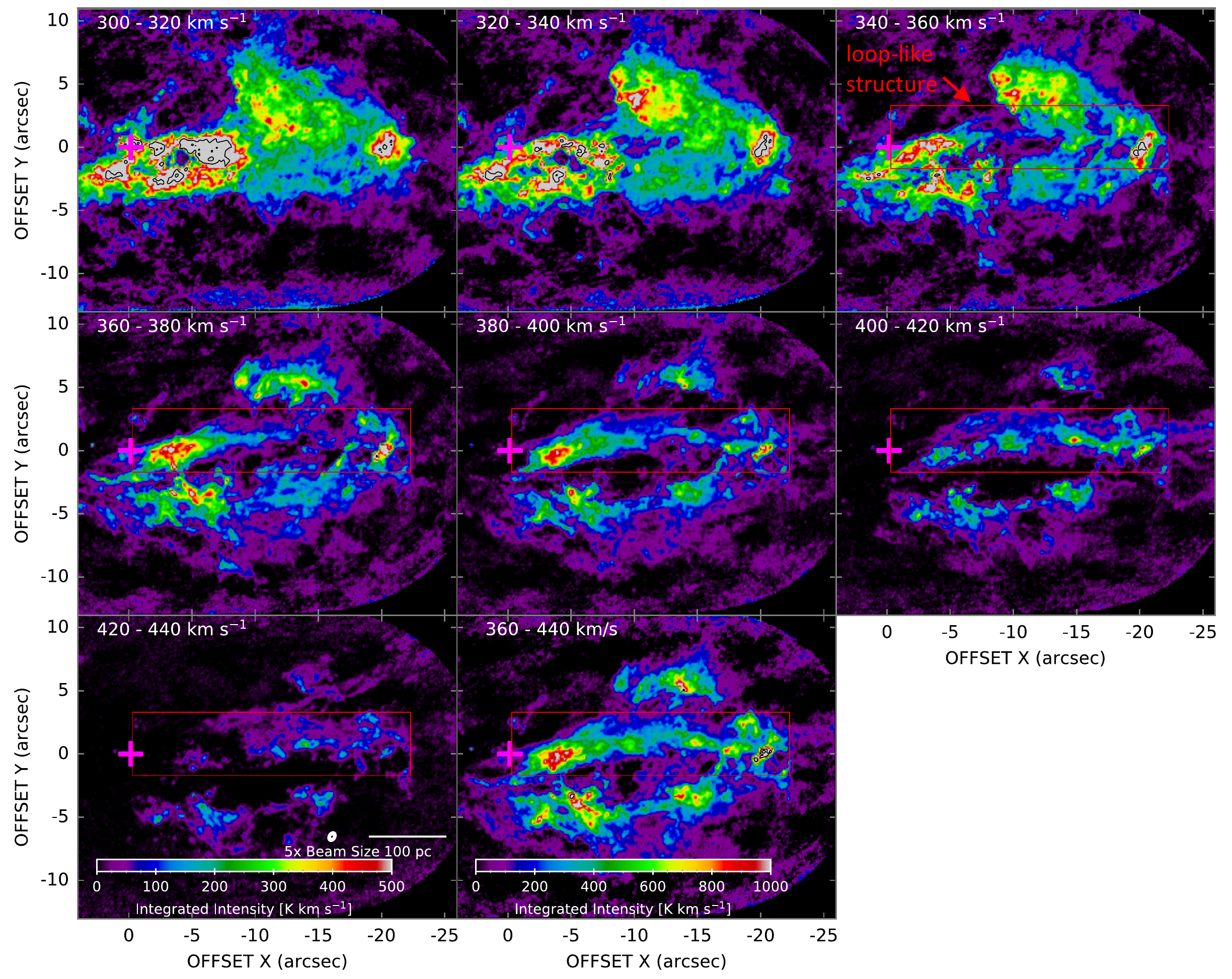}
\caption{Velocity channel map of $\cott$.
Contours are shown at 1200 and 1600 K $\kms$ for the bottom-center (last) panel and 700 and 1100 K $\kms$ for the other panels.
A pink cross indicates the position of the center of NGC~253 \citep[][]{mul10}.
A red rectangle corresponding the red rectangle in Figure~\ref{fig:fig_chn_wholedisk} is the region including the loop-like structure.
The bottom-center panel is the same as Figure~\ref{fig:fig_loop_rot35}a.
\label{fig:fig_loop32_grid}}
\end{figure}

Figure~\ref{fig:fig_loop32_grid} shows detailed velocity channel distribution for the loop-like structure integrating every 20 $\kms$.
The bottom rightmost panel is the same image as Figure~\ref{fig:fig_loop_rot35}a.
In ($-$10$\arcsec$ $\lesssim$ OFFSET $X$ $\lesssim$ $+$3$\arcsec$, $-$5$\arcsec$ $\lesssim$ OFFSET $Y$ $\lesssim$ $+$2, $+$340 $\kms$ $\lesssim$ $\vl$ $\lesssim$ $+$360 $\kms$), CO emissions from the western base of the loop-like structure and the disk component are contaminated.
At $\vl$ = 360 $\kms$ the middle region gas of the loop-like structure emerges and moves to the negative OFFSET $X$ as the velocity increase until $\vl$ = 440 $\kms$.
The every 20 $\kms$ integrated intensity of the bases of the loop-like structure shows $\sim$500 K $\kms$ at most, which corresponds to 2.5 times higher than the middle region.

\begin{figure}[htbp]	
\centering
\includegraphics[width=90mm]{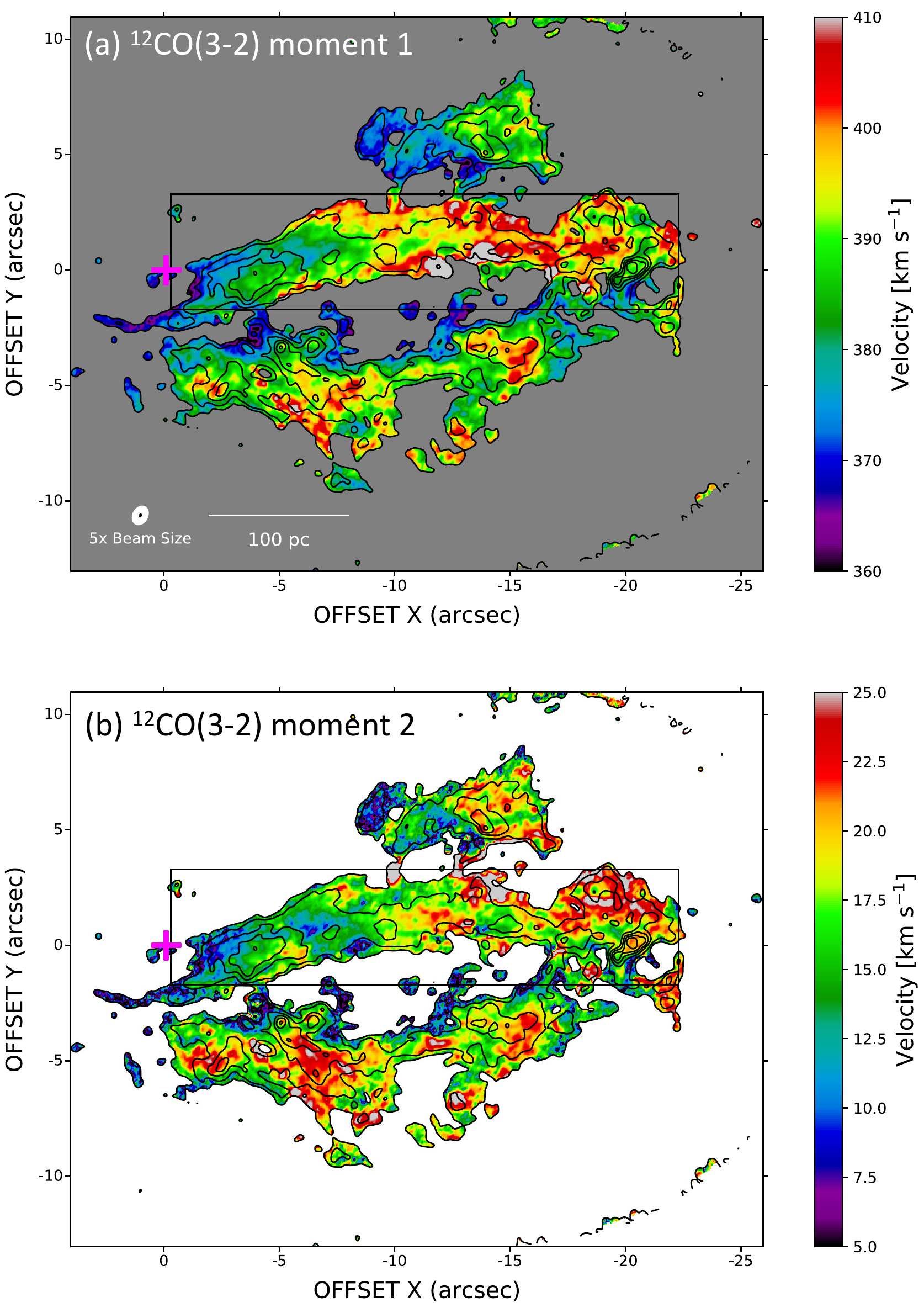}
\caption{(a) Distribution of intensity-weighted mean velocity (moment 1) of $\cott$ integrated from $\vl$ = 360 to 440 $\kms$.
Voxels with the intensity below 10 $\sigma$ (1$\sigma$=13 K $\kms$) are masked in advance.
Contours are plotted at 130, 260, 520, 780, 1040, and 1300 K $\kms$.
A pink cross indicates the position of the center of NGC~253.
(b) Distribution of the square root of the intensity-weighted velocity dispersion (moment 2) of $\cott$ integrated from $\vl$ = 360--440 $\kms$.
Contours and the pink cross are same as Figure~\ref{fig:fig_loop32_rot35_mom1_2}a.
\label{fig:fig_loop32_rot35_mom1_2}}
\end{figure}

Figures~\ref{fig:fig_loop32_rot35_mom1_2}a and \ref{fig:fig_loop32_rot35_mom1_2}b show intensity-weighted mean velocity distribution (moment 1) and the square root of the intensity-weighted velocity dispersion distribution (moment 2) of $\cott$ emission in the velocity range of $\vl$ = 360 to 440 $\kms$, respectively.
Figure~\ref{fig:fig_loop32_rot35_mom1_2}a again exhibits the velocity trend, linearly increase along with the middle region from east to west, from $\vl$ = 370 to 420 $\kms$.
Also, we find the velocity increase along with the width of the loop-like structure from the inner part to the outer part.
The western base shows slightly lower velocity of $\vl$ = 390 $\kms$ than the middle region because the western base region is extending in velocity and has a peak at the lower velocity than the middle region. (see Figure~\ref{fig:fig_CO32_pv}).

Figure~\ref{fig:fig_loop32_rot35_mom1_2}b suggests that the velocity dispersion of the loop-like structure does not have a significant variation and it is typically 10--20 $\kms$.
The small values toward the bases are caused by the velocity range of the moment maps, which focuses on the velocity of the middle region and only partly includes the bases.

\subsection{$\cott$/$\cooz$ ratios}\label{result:ratio}

Here we investigate excitation states in the giant molecular loop-like structure by using the line-intensity ratio between $\cott$ and $\cooz$.
First we smoothed the $\cott$ data set to the beam size of 1$\farcs$6 $\times$ 1$\farcs$2 to match that of the $\cooz$ data set and then regridded both the $\cooz$ and smoothed $\cott$ data sets to 0$\farcs$3 and 5 $\kms$.
We next masked voxels under 10$\sigma$, which corresponds to 2 K $\kms$ for $\cooz$ and 0.3 K $\kms$ for $\cott$.
Finally we integrated these data sets every 20 $\kms$ from $\vl$ = 360 to 440 $\kms$ and made a ratio map.

\begin{figure}[htbp]	
\centering
\includegraphics[width=160mm]{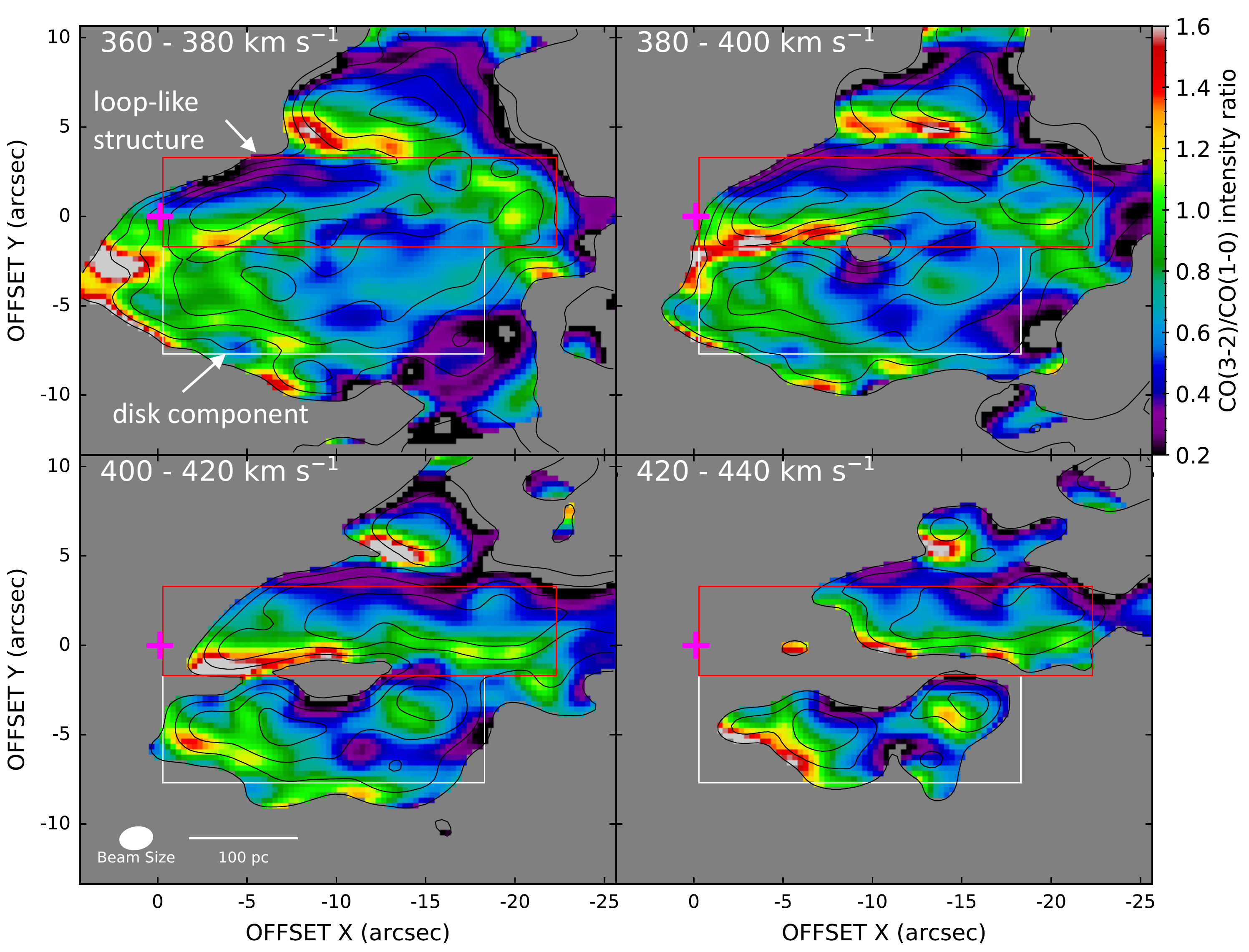}
\caption{Velocity channel distribution of the intensity ratio between $\cott$ and $\cooz$.
Pixels under 10$\sigma$ are masked in advance for $\cooz$ (1$\sigma \sim$2 K $\kms$) and for $\cott$ (1$\sigma \sim$0.3 K $\kms$).
Contours are drawn at 40, 80, 160, and 320 K $\kms$.
A pink cross indicates the position of the center of NGC~253 \citep[][]{mul10}.
The red rectangle is the region including the loop-like structure.
\label{fig:fig_loop_32over10_ratio}}
\end{figure}

Figure~\ref{fig:fig_loop_32over10_ratio} shows velocity channel distribution of the intensity ratio between $\cott$ and $\cooz$ (hereafter $\ratio$) toward the molecular loop-like structure.
The disk component indicated by the white box has the typical value of $\ratio$ of 0.6 and most pixels of the component display $\ratio$ under 1.3.
On the other hand, the eastern and western bases of the loop-like structure have significantly higher values of 0.8--1.6 and $\sim$ 1.2 in the velocity range of 360--420 $\kms$ and 360--400 $\kms$, respectively.
The middle region of the loop-like structure shows $\ratio$ of 0.4--1.4 in the velocity range of 380--400 $\kms$ and there is significant difference of ratios between the inner part (0.8 to 1.4) and the outer part (0.4 to 0.8) of the middle region.
The averaged value of $\ratio$ in the loop-like structure is 0.69.
From these results, we found that $\ratio$ is not so different between in the loop-like structure and disk component but the bases and inner part of the loop-like structure have significant higher values.

\subsection{Mass of the loop-like structure}\label{result:mass}

We here derive mass of the loop-like structure from the integrated intensity of $\cooz$ emission.
The following equation gives molecular column density $\nhtwo$.
\begin{equation}
\label{eq:xfactor}
\nhtwo = X_{\text{CO}} I_{\text{CO}}.
\end{equation}

Here $X_{\text{CO}}$ and $I_{\text{CO}}$ are the $X$ factor, which is an empirical conversion factor to estimate $\nhtwo$ and the integrated intensity of $\cooz$, respectively.
We used the $X$ factor of 0.5 $\times$ 10$^{20}$ [K $\kms$ cm$^{-2}$]$^{-1}$, which is the value often used for the CMZ of NGC~253 \citep{ler15}.
The above $X$ factor corresponds to $\alpha_{CO}$ of 1.1 $\sm$ [K $\kms$ pc$^{-2}$]$^{-1}$, which is the conversion factor from molecular mass to integrated intensity.
The molecular mass, $M$ is derived by
\begin{equation}
\label{eq:alphafactor}
M = \alpha_{CO} L_{\text{CO}}.
\end{equation}

Here $L_{\text{CO}}$ is the luminosity of $\cooz$.
The molecular mass derived from inside the red rectangle in Figure~\ref{fig:fig_loop_rot35}b is 1.3 $\times$ 10$^{7}$ $\sm$.
We defined areas for the bases as the $\cott$ FWHM intensity.
The derived masses are 5.9 $\times$ 10$^{5}$ $\sm$, 1.7 $\times$ 10$^{5}$ $\sm$ for the eastern and western bases and the mass ratio between the loop-like structure's whole mass and the total mass of the bases is 5$\%$.

Next we derived $\nhtwo$ from the higher resolution $\cott$ data set by applying $\cott$/$\cooz$ =0.69.
Note that pixels with $\le$10$\sigma$ (=130 K $\kms$) in the red rectangle of this data set are also masked in advance.
The derived maximum molecular column densities in the eastern base, western base, and the middle region are 8.1 $\times$ 10$^{22}$ cm$^{-2}$, 1.3 $\times$ 10$^{23}$ cm$^{-2}$, and 5.8 $\times$ 10$^{22}$ cm$^{-2}$, respectively.
These physical parameters are summarized in Table~\ref{tab:loop_mass}.

\begin{deluxetable*}{cccc}
\tabletypesize{\scriptsize}
\tablewidth{0pt}
\tablecaption{Molecular Mass and Maximum Gas Column Density in Eastern Base, Western Base and Middle Region
\label{tab:loop_mass}}
\tablehead{
\colhead{}
 & Eastern Base & Western Base & Middle Region
\\[-2.5mm]
}
\startdata
Molecular mass [$\sm$] & 1.16 $\times$ 10$^{6}$ & 1.53 $\times$ 10$^{5}$ & 1.17 $\times$ 10$^{7}$ \\
Maximum gas density [cm$^{-2}$] & 8.1 $\times$ 10$^{22}$ & 1.3 $\times$ 10$^{23}$ & 5.8 $\times$ 10$^{22}$ \\
\enddata
\end{deluxetable*}

\subsection{Optical depth of the bases}\label{discussion:opt_dep}
It is possible that the $\ratio$ at the two bases of the loop-like structure is overestimated due to their higher optical depths than the middle region.
Therefore, we here estimate the optical depths of $\cooz$ and $\cott$ at the bases.
we adopted the velocity range of 360--440 $\kms$ for the bases.
The optical depth $\tau$ in the local thermodynamic equilibrium (LTE) condition is given by
\begin{equation}
\label{eq:opt_dep}
\tau = \frac{4\pi^{3}N\nu\mu^{2}}{3kT\Delta V}\exp(-\frac{Jh\nu}{2kT})(1-\exp(-\frac{h\nu}{kT}))
\end{equation}
where $\nu$, $N$, $\mu$, $k$, $h$, $T$, $J$, and $\Delta V$ are the rest frequency, column density of molecular gas, electric dipole moment, Boltzmann constant, Planck constant, kinetic temperature, lower-rotational level, and velocity dispersion, respectively.
We here used molecular column density of $\sim$8.1$\times$10$^{22}$ cm$^{-2}$ and $\sim$1.3$\times$10$^{23}$ cm$^{-2}$, and the velocity dispersions of $\Delta V\sim$50 $\kms$ and $\sim$70 $\kms$, for the eastern and western bases, respectively.
Assuming the kinematic temperature of 35 K \citep[][]{ler15} and CO/H$_{2}$ of 10$^{-4}$, the optical depths of eastern and western bases are calculated to be $\sim$2.9 in $\cooz$, $\sim$14 in $\cott$, and $\sim$3.3 in $\cooz$, $\sim$16 in $\cott$, respectively.
These results show that the optical depths of $\cooz$ and $\cott$ are high enough at the both the bases and thus the possibility of under/over estimation of $\ratio$ may be low enough.


\section{Discussion}\label{sec:dis}

Stars and the interstellar medium are strongly concentrated within the central region of galaxies.
This concentration makes the gravitational potential well deepest in a galaxy, attracting gas into the region \citep[][]{mor96}.
This gas attraction leads to active star formation, which accelerates the gas motion by the stellar feedback \citep[e.g.,][]{bol13}.
Such gas motion may also be accelerated via Parker instability \citep[][]{par66} as observationally confirmed by the magnetically floated loops.
As such, the gravitational potential well may be controlling gas motion via stellar feedback and Parker instability, and it is important to understand the origin of the gas kinematics in the galactic central region in our efforts to elucidate galactic evolution. 

\subsection{Origin of the non-circular motion}\label{discussion:ori_noncir}

As shown in Section~\ref{result:distribution}, in the present work, we discovered a loop-like structure whose velocity is $\vl$ = 360 to 440 $\kms$ from the CMZ of NGC~253.
We interpret that the loop-like structure indicates a non-circular component whose velocity is deviated from the disk velocity by 50 $\kms$.
\citet[][]{kri19} identified a "nondisk" component that does not follow the velocity field of the disk or bar \citep[][; Figure 6]{kri19}, which indicates the loop-like structure is a "nondisk" component.
In the central region of the MW, the origin of the gas with non-circular motion is usually interpreted in terms of the radial motion driven by the bar potential \citep[e.g.,][]{bin91, rod06}.
In the MW, the bar is in the line of sight toward the plane and delineates a parallelogram within $\pm$ 250 $\kms$~in the p-v diagram. 
NGC~253 is classified as SAB(s)c \citep[][]{deV91} and the bar potential of NGC~253 is supposed to be weaker than in the MW classified as SBc.
However, the difference between the two bars are not well known quantitatively due to the limited sensitivity of near-infrared/optical observations in NGC~253.
Future observations and simulations are desirable to clarify the issue.
Here we assumed that the NGC~253 bar has a potential well comparable to that in the MW, and the NGC~253 bar is accelerating the gas comparable to that in the MW.
Considering that angle between the line of sight and the bar major axis is 16$^{\circ}$, which is almost parallel, the bar acceleration reaches $\sim$250 $\kms$ toward the bar direction in the MW \citep[][]{bin91}.
The angle between the major axis and the NGC~253 bar is 17$^{\circ}$ \citep[][]{sco85}, i.e., the line of sight makes an angle of 73$^{\circ}$ to the bar and, considering the inclination of $\sim$75$^{\circ}$ \citep[][]{kun07}, the projected velocity to the line of sight is estimated to be 250 $\kms$ $\times$ cos(73$^{\circ}$) $\times$ sin(75$^{\circ}$) $\sim$71 $\kms$ in the NGC~253, about 0.3 times of that of the MW.
It is therefore unlikely that the loop-like structure is accelerated by the bar.
On the other hand, \citet[][]{kri19} argued that the "otherdisk" in the "nondisk" needs to be explained by some additional kinematic component of the disk/ bar.
If we assume that there is a small bar with a size similar to the CMZ of NGC~253, the gravitational potential of the bar may explain the large velocity offset of at most 200 $\kms$ from the disk component observed in the loop-like structure.
Future hydrodynamic simulations with a such bar is needed for testing the above assumption.
However, it might be considerably difficult to explain the warp in the middle region of the loop-like structure only by the bar potential.
Thus, we need more plausible scenario that can explain the lifted-up structure against the gravity.

\subsection{Origin of the velocity distribution of the loop-like structure}\label{discussion:ori_vel}

As discussed in Section~\ref{discussion:ori_noncir} the NGC~253 loop-like structure cannot be understood in terms of the bar potential, and some other driving mechanism is required.
It has been well established in the literature that there is starburst and a number of high-mass stars are being formed in the CMZ of NGC~253 \citep[e.g.,][]{kkor09, ben15, ler15, ler18, dav16, and17, man19, coh20, ric20, lev21, mil21}.

Observations indeed suggest that supershells driven by supernova explosions \citep[e.g.,][]{sak06} and outflows by starburst \citep[e.g.,][]{bol13}.
In the following, we explore a possible role of the supershell model as the origin of the loop-like structure.

\subsubsection{Supershell model}\label{discussion:supershell}

\begin{figure}[p]		
\centering
\includegraphics[width=150mm]{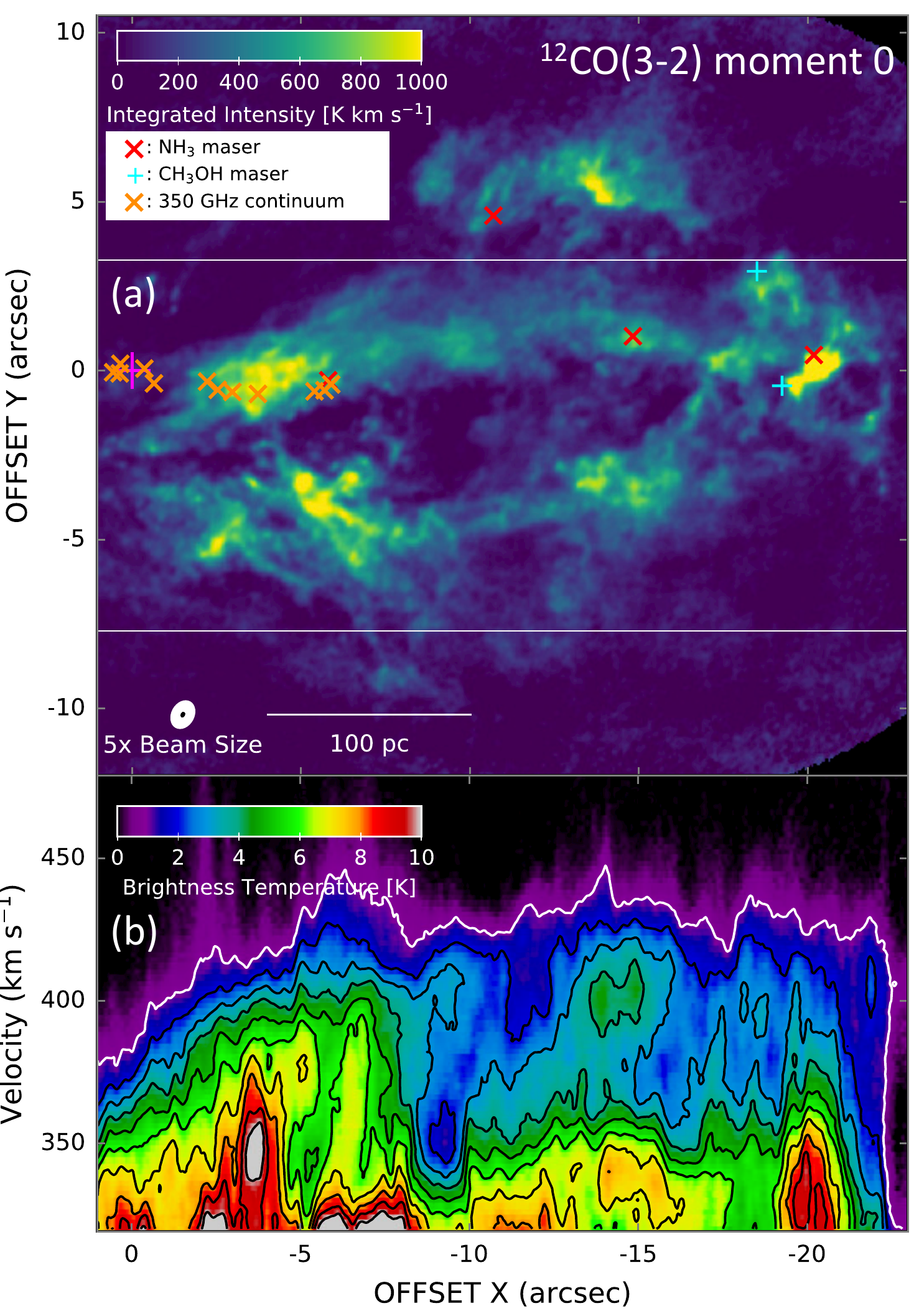}
\caption{(a) Integrated intensity map of $\cott$ integrated from $\vl$ = 360 to 440 $\kms$. Positions of the 350 GHz continuum peaks \citep{ler18}, class $\text{\sc{I}}$ CH$_{3}$OH masers \citep{gor17}, NH$_{3}$ masers \citep{gor17} are plotted as orange x symbols, blue crosses, and red crosses, respectively. The pink cross indicates the position of the center of NGC~253. Two solid lines indicate the integrated OFFSET $Y$ range for Figure~\ref{fig:fig_loop32_vs_SSC}. (b) Position-velocity diagram of $\cott$ integrating the whole the shell width of 187 pc. Contours are plotted at 1, 2, 3, 4, 5, 6, 7, 8, 9, and 10 K.
\label{fig:fig_loop32_vs_SSC}}
\end{figure}

In the model, the ambient gas is accelerated radially by the explosive winds and we expect an elliptical pattern of the accelerated gas in the p-v diagram.
Figure~\ref{fig:fig_CO32_pv}a shows that the loop-like structure has a uniform velocity gradient as well as large velocity dispersion and high brightness in the loop-like structure bases.
This is quite different from what is expected in a SNe accelerated shell.
If we assume that the cavity in $-$16$\arcsec$ $\lesssim$OFFSET $X$ $\lesssim$ $-$8$\arcsec$, $\vl$ = 360--400 $\kms$ was formed by an SNe we expect that there is a counter part of the shell at velocity less than $\vl$ = 360 $\kms$.
We, however, do not confirm the low velocity counterpart of the shell at velocity less than $\vl$ = 340 $\kms$ (see Figure~\ref{fig:chn_ap2} in Appendix).
It is possible that the cavity in the p-v diagram in Figure~\ref{fig:fig_CO32_pv}a is a feature formed by chance via overlapping of the loop-like structure and the part of disk component.
Figure~\ref{fig:fig_loop32_vs_SSC}a shows an overlay of the loop-like structure with the 350 GHz continuum \citep[][]{ler18}, NH$_{3}$ maser \citep[][]{gor17}, and class $\text{\sc{I}}$ CH$_{3}$OH maser \citep[][]{gor17}.
This indicates that a combination of the loop-like structure and the part of the disk component looks like a shell-forming cavity centered at OFFSET $X$ = $-$11$\arcsec$, OFFSET $Y$ = $-$2$\farcs$2.
Accordingly, if we consider that the loop-like structure and the disk component form a shell, we expect that a cluster is located within the shell.
Contrary to the expectation, the cluster is located at the base of the loop-like structure and there is no cluster in the shell.
This suggests that the cavity was not formed by the cluster.
Figure~\ref{fig:fig_loop32_vs_SSC}b shows the loop-like structure and the disk component in a p-v diagram, indicating no cavity corresponding to the shell. We are thus forced to interpret that the shell is only a half in the north, where the other half disappeared by some unknown cause.
If we assume this scenario, the expansion velocity of the shell is 80 $\kms$ and the kinetic energy of the shell becomes $\sim$8.3 $\times$ 10$^{53}$ erg.
Considering the total kinetic energy release of an SNe, the energy is too large to be supplied by a single SNe.
If we assume that all the kinetic energy comes from SNe, 8300 SNe, or a cluster of 10$^{6--7}$ $\sm$, is required in total under an assumption of 10 \% conversion efficiency.
There is no observational sign of such a cluster in the cavity.
The expansion velocity of a super shell is 20 $\kms$ \citep[][]{fuk99, kmat01}, much smaller than that observed in the present feature 80 $\kms$.
Only the eastern base of the loop-like structure where four proto super star cluster (hereafter proto-SSC) candidates \citep[\#4-7; ][]{ler18} are concentrated is the region significantly affected by star forming feedbacks.
From the molecular mass (1.16 $\times$ 10$^{6}$ $\sm$) and velocity width (80 $\kms$) of the eastern base, the kinetic energy is estimated to be $\sim$7.4 $\times$ 10$^{52}$ erg.
On the other hand, the outflow energy of the most energetic proto-SSC candidate ($\#$5) is $\sim$4 $\times$ 10$^{50}$ erg \citep[][]{lev21}.
Thus, given 10 $\%$ of the energy is injected into the surrounding gas, the estimated total kinematic energy gain by outflows is less than $\sim$1.6 $\times$ 10$^{50}$ erg.
This reaches only 0.22\% of the kinematic energy of the eastern base, thus not enough to accelerate gas in the region.
Therefore, the outflows from the proto-SSC candidates cannot explain the energy of the loop-like structure.

The above discussion suggests that the velocity and energetics of the present loop-like structure are difficult to understand in terms of SNe, and some other mechanism needs to be considered.
As such mechanisms in the MW, we find two possibilities in the previous works; one is the intermediate mass black hole (IMBH) of 10$^{4-5}$ $\sm$ \citep[e.g.,][]{oka16} and the other the magnetic flotation loop-like structure driven by the Parker instability \citep[e.g.,][]{fuk06}.

\subsubsection{IMBH model}\label{discussion:imbh}

An IMBH gravitationally attracts the ambient gas, which circulates in the Kepler orbit around the IMBH and creates molecular gas having large velocity dispersion.
\citet[][]{tak19} proposed that the molecular gas around an IMBH having $\sim$10$^{5}$ $\sm$ mass has mass of $\sim$10$^{4}$ $\sm$, with the radius of $\sim$5 pc, showing velocity width of 100 $\kms$.
So if there is a group of IMBHs, a high-velocity, larger-sized cloud is represented.
The present loop-like structure has total mass of $\sim$10$^{7}$ $\sm$, suggesting that it may include 1000 IMBHs with the mass of $\sim$10$^{5}$ $\sm$, whereas their total mass amounts of IMBHs to $\sim$10$^{8}$ $\sm$.
The length and width of the loop-like structure are 300 pc and 50 pc, respectively, thus 60 $\times$ 10 IMBHs are required.
However, such a large number and distribution of IMBHs is quite unnatural.
We consider that the scenario is unlikely.
It is unlikely that such spatial distribution takes place locally in the CMZ of NGC~253, and we do not consider the possibility in the following.

\subsubsection{MHD model}\label{discussion:mhd}
In the molecular gas, which is in dynamical equilibrium, the gravitational force is in equilibrium with the magnetic pressure.
This configuration is, however, unstable and the magnetic field tends to rise up against gravity once the equilibrium breaks down locally to form the loop-like gas distribution.
This rising motion is time dependent and is known as the Parker instability \citep[][]{par66}.
On the solar surface, the instability has been observed over several decades.
In the interstellar space, at a typical size of 10--100 pc the Parker instability was discovered in the central region of the MW as loops 1 and 2 by \citet[][]{fuk06}.
\citet[][]{mat88} carried out two-dimensional MHD numerical simulations of the galactic disk and showed that the magnetic loop grows and the gas falls down along the loop supersonically to create a shocked region.
\citet[][]{mac09} extended the simulations to a global 3D configuration and showed that more than 400 loops are formed in a 1 kpc disk.
\citet[][]{suz15} developed refined 3D simulations and demonstrated the parallelogram pattern in the p-v diagram is reproduced only by the MHD effect without bar potential.
These theoretical works lend support for the idea that MHD effect provides strong mechanism of agitating gas in the magnetized gas disk.
NGC~253 is a place where the picture is applicable along with the CMZ.

\subsection{Magnetic loop scenario}\label{discussion:sce}

In Section~\ref{discussion:mhd}, we recognized that the loop-like structure in NGC~253 has a shape similar to a magnetically floated loop as theoretically predicted.
We will test if this is a viable mechanism in NGC~253 and explore its implications.

\subsubsection{Plausibility of the magnetic loop}\label{discussion:plau_mhd}

In the formation of a magnetically floated loop, a strong magnetic field that is energetic enough for the flotation in the vertical direction to the disk is required.
We estimate the field strength required for the NGC~253 loop-like structure.
By assuming that the kinetic energy of the molecular gas is equal to the magnetic field energy, the following relation is used where $\rho$ [kg~m$^{-3}$] is gas density, $V$ [m~s$^{-1}$] is the turbulence velocity, $B$ [T] is magnetic field strength, and $\mu _{0}$ = 4$\pi$ $\times$ 10$^{-7}$ N~A$^{-2}$ is the vacuum magnetic permeability,

\begin{equation}
\label{eq:magpre}
\frac{1}{2}\rho V^{2} = \frac{B^{2}}{2\mu _{0}}
\end{equation}

Assuming the shape of the NGC~253 loop-like structure is a cylinder with bases corresponding to the eastern and western bases, we estimate the gas density of the NGC~253 loop-like structure of $n$ = 320 cm$^{-3}$ adopting the gas mass of 1.3 $\times$ 10$^{7}$ $\sm$, the loop-like structure width of 50 pc, and the loop-like structure length of 300 pc.
The density is combined with the average velocity dispersion over the loop-like structure as $V$ = 43 $\kms$ and the field strength of $\sim$0.5 mG for producing the NGC~253 loop-like structure.
This is a factor of three times larger than that required for loops 1--3 in the MW.
If we adopt field strength of 1--10 mG as found in the MW \citep[e.g.,][]{mor96}, the present scenario seems to be reasonable in NGC~253.
The loop-like structure height in NGC~253 is 20 pc, while the typical loop-like structure height in the MW is 100 pc \citep[][]{fuk06}.
The loops 1--3 are located at distance of 500--1000 pc from the Galactic center where the gravitational field is weaker than the inner part within 100--300 pc.
The scale height of a magnetic loop is inversely proportional to the gravitational potential or, in other words, stellar density \citep[][]{ktak09}.
According to \citet[][]{pmat17}, the stellar density within the central $\sim$200 pc is a few $\times$10$^{10}$ $\sm$ kpc$^{-2}$, thus the gravitational potential expected at the loop-like structure is one order of magnitude higher than that of loops 1--3 in the Galactic center.
The scale height at the position of the loop-like structure then must be 10 times lower than that of loops 1--3, which corresponds to $\sim$20 pc and the lower height of the loop-like structure does not conflict with the magnetic loop scenario.

We infer that the stronger gravitational force leads to the reduced scale height in the NGC~253 loop-like structure.

\subsubsection{Comparison with magnetic loops in the MW {\sc I}: Physical properties of the magnetic loops}\label{discussion:comp_mw1}

We shall compare the physical parameters of the NGC~253 loop-like structure (hereafter N253 loop 1) with those in the MW loops.
In the MW, there are five loops/loop candidates in the previous papers; the loops 1, 2, Bridging component for the broad velocity features at $l$ $\sim$ 3$^{\circ}$--5$^{\circ}$ (hereafter $l$ $\sim$ 3$^{\circ}$--5$^{\circ}$ loop); \citep[][]{fuk06}, the loop 3; \citep[][]{fuj09}, and the loop connecting Sgr~A and Sgr~B2; \citep[][]{eno21_2}.

The loops 1 and 2 having negative velocities are located at 500--600 pc from the Galactic center, and are connected at their footpoints.
The loop 3 is also located at 500 pc from the Galactic center and has positive velocity.
The $l$ $\sim$ 3$^{\circ}$--5$^{\circ}$ loop is located at 700 pc from the Galactic Center in projection while the loop connecting Sgr~A and Sgr~B2 is 100 pc from the Galactic Center in projection and all of them have positive velocities.

Table~\ref{tab:table_loop_phy} lists the physical parameters of the four loops in the MW along with the present the N253 loop 1.

\begin{deluxetable*}{cccccccc}	
\tabletypesize{\scriptsize}
\tablewidth{0pt}
\tablecaption{Physical properties of the loop in NGC~253 and the loops 1--3 in the MW 
\label{tab:table_loop_phy}}
\tablehead{
\colhead{}
(1) & & N253 loop 1 & loop 1 & loop 2 & loop 3 & $l$ $\sim$ 3$^{\circ}$--5$^{\circ}$ loop & Sgr~A and Sgr~B2
\\[-2.5mm]
}
\startdata
(2) & Distance [pc] & 200 & 500 & 600 & 500 & 700 & 100 \\
(3) & Length [pc] & 300 & 500 & 300 & 600 & 400 & 100 \\
(4) & Width [pc] & 50 & 30 & 30 & 70-150 & 50 &  \\
(5) & Height [pc] & 20 & 220 & 300 & 300 & 100 & 20 \\
(6) & $\vl$ [$\kms$] & $+$360--$+$440 & $-$180--$-$90 & $-$90--$-$40 & $+$30--$+$160 & $+$10--$+$100 & $+$0--$+$80 \\
(7) & Molecular mass [$\sm$] & 1.3$\times$10$^{7}$ & $\sim$10$^{5}$ & $\sim$10$^{5}$ & 3$\times$10$^{6}$ &  &  \\
(8) & Time scale [yr] & 10$^{6}$ & several$\times$10$^{6}$ & several$\times$10$^{6}$ & 10$^{7}$ &  & \\
(9) & Mass ratio & 10\% & 41\% & 38\% & 16\% &  & \\
(10) & Reference & 1 & 2 & 2 & 3 & 2 & 4 \\
\enddata
\tablecomments{Rows: (1) names, (2) distances from the center of the hosted galaxies, (3) lengths of loops, (4) widths of loops, (5) heights of loops from the galactic plane (the footpoints), (6) velocity ranges for loops in LSR velocity, (7) total molecular masses of loops, (8) estimated dynamical time scales of loops, (9) molecular mass ratios between footpoint masses and whole the loop masses, (10) references (1, this study; 2, \citet[][]{fuk06}; 3, \citet[][]{fuj09}; 4, \citet[][]{eno21_2}})
\end{deluxetable*}

\subsubsection{Comparison with magnetic loops in the MW {\sc II}: The spatial distribution}\label{discussion:comp_mw2}

The N253 loop 1 is separated from the galactic center by $\sim$200 pc, which is closer than the loops 1--3 and $l$ $\sim$ 3$^{\circ}$--5$^{\circ}$ loop at a distance of $\sim$700 pc from the center of the MW and farther than the loop connecting Sgr~A and Sgr~B2 at a distance of $\sim$100 pc from the center of the MW.
The N253 loop 1 has length of $\sim$300 pc and width of $\sim$50 pc, while the average lengths of the loops 1--3, the $l$ $\sim$ 3$^{\circ}$--5$^{\circ}$ loop, and the loop connecting Sgr~A and Sgr~B2 are 400 pc and the widths are 60 pc; these values are similar to those of the N253 loop 1.



The height of the loops seems to be different between NGC~253 and the MW.
The height of the N253 loop 1 is $\sim$20 pc, an order of magnitude smaller than that of the loops 1--3 and $l$ $\sim$ 3$^{\circ}$--5$^{\circ}$ loop $\sim$100--200 pc.
The low height is connected with the gravity toward the plane, which decreases with radius from the galactic center \citep[e.g.,][]{mor96}.
Also, as we discussed in Section~\ref{discussion:plau_mhd}, the scale height of the N253 loop 1 is estimated to be $\sim$20 pc, and hence the height does not contradict the magnetic loop scenario.
On the other hand, the height of the N253 loop 1 is almost similar to that of the loop connecting Sgr~A and Sgr~B2 ($\sim$20 pc).
The quite close projected distances from the galactic center for these two loops, ---100 pc for the loop connecting Sgr~A and Sgr~B2 and 200 pc for the N253 loop 1---might be attributed to the small height of the loops.
If we assume the Alfven speed to be $\sim$24 $\kms$ \citep[][]{fuk06}, the time scale of the N253 loop 1 is estimated to be $\sim$1 Myr because a loop grows at the Alfven speed \citep[][]{mat88}.
This suggests that the N253 loop 1 is several times younger than loops 1--3. The CO intensity is higher in the footpoints than in the bridge. The intensity ratio between the footpoint and the bridge in the loops 1--3 is 2$\sim$4:1 in $\cooz$, while that in NGC~253 is 1.4:1, suggesting that the N253 loop 1 has a higher gas fraction in the bridge than in the footpoint as compared with the loops 1--3. This is consistent with the time scale difference which may represent the younger age of the N253 loop 1.
By the magnetic floatation, the gas coupled with the field lines is lifted up away from the disk and the gas falls down along the field lines to the footpoint.
This suggests that the younger loop tends to have more gas in the bridge than in the footpoint.
The maximum molecular column density of the footpoint of the N253 loop 1 is $\sim$10$^{23}$ cm$^{-2}$.
The maximum column density of the footpoint in the loops 1--3 is 10$^{22}$ cm$^{-2}$ \citep[][]{eno21}, 10 times less than in the N253 loop 1.
The total gas mass in the N253 loop 1 is 10$^{7}$ $\sm$, 100 times more than in the loops 1--2 and 10 times more than in the loop 3.
The cause of the larger mass in the N253 loop 1 than in the loops 1--3 may be ascribed to the large ambient gas density in the CMZ of NGC~253.
The loops 1--3 region has gas column density of 10$^{22}$cm$^{-2}$ \citep[][]{eno21}, while that in the CMZ of NGC~253 is 50 times larger, i.e. $\sim$5 $\times$ 10$^{23}$ cm$^{-2}$ \citep[][]{sak11}.

\subsubsection{Comparison with magnetic loops in the MW {\sc III}: The velocity distribution}\label{discussion:comp_mw3}

Next, we compare the velocity distribution among the relevant features.
The N253 loop 1 has velocity width of $\sim$50--100 $\kms$ in the footpoints with a velocity gradient of 40 $\kms$ vertical to the loop.
The linewidths of the loops 1--3, the $l$ $\sim$ 3$^{\circ}$--5$^{\circ}$ loop and the loop connecting Sgr~A and Sgr~B2 are typically 50--100 $\kms$.
The line width of the N253 loop 1 is one to two times smaller than that of the $l$ $\sim$ 3$^{\circ}$--5$^{\circ}$ loop, while the N253 loop 1 is almost similar to those of the loops 1--3, and the loop connecting Sgr~A and Sgr~B2.
The velocity span of the loops 1--3, the $l$ $\sim$ 3$^{\circ}$--5$^{\circ}$ loop, and the loop connecting Sgr~A and Sgr~B2 are 90, 50, 130, 150, and 80 $\kms$, respectively.
The average of velocity span is $\sim$80 $\kms$ and this value is close to that of the N253 loop 1. 

This suggests that the velocity distribution of the N253 loop 1 looks similar to the loops 1--3.

On the other hand, the velocity decrease toward the center of the loop's width observed in the N253 loop 1 is not found in the magnetic loops in the MW \citep[loops 1, 2;][]{tor10_2}.
It is possible that this decrease is accompanied by an overlap of thin loops.
\citet[][]{mac09} performed 3D MHD simulations and suggested that a magnetic loop is often formed as a bundle of thin magnetic loops.
We then assume that the N253 loop 1 (main loop), which has a width of 50 pc, overlaps with a thinner loop (hereafter a sub loop) with a width of $\sim$20 pc.
The bridge of the sub loop is seen at $\vl$ = 360--380 $\kms$ in Figure~\ref{fig:fig_loop32_grid} while that of the main loop is seen at $\vl$ = 380--400 $\kms$.
We guess the sub loop, which has lower velocity and lower maximum height than the main loop, overlaps at the center of the main loop's width and is bundled with the main loop, and this causes the centroid velocity decrease at the center of the N253 loop 1 seen in Figure~\ref{fig:fig_loop32_rot35_mom1_2}.

\subsubsection{Comparison with magnetic loops in the MW {\sc IV}: The spatial distribution of gas excitation}\label{discussion:comp_mw4}

Next we test the excitation states of gas.
The $\ratio$ in the footpoint and bridge of the N253 loop 1 are 0.8--1.4 at $\sim$30 pc resolution, and these are higher than that in the disk component (0.6).
This indicates that some extra excitation mechanism, not working in the disk, is needed in the loop.
The trend that the $\ratio$ is enhanced in the footpoint and the inner part of the N253 loop 1 is similar to with those in the loops 1, and 2 \citep[][]{kud11}.
\citet[][]{tor10,kud11} suggest that the origins of the high-excitation gas at footpoints are shock, gas compression, or gas heating.
This suggests that the enhanced excitation condition of the N253 loop 1 may be ascribed to the shock or gas compression at the footpoints due to magnetic flotation.

\subsubsection{New insights of the magnetic loop}\label{discussion:new_insights}

In Section~\ref{discussion:comp_mw1}--\ref{discussion:comp_mw4}, it was shown that the N253 loop 1 has properties different from those of the loops 1--3.
First, because the CMZ of NGC~253 is more massive than the central region of the MW, the gas mass of the N253 loop 1 is 10--100 times more massive than in the loops 1--3.
Further, the molecular mass ratio between footpoint masses and the entire loop masses of the N253 loop 1 is approximately two to four times lower than that of the loops 1--3  and the N253 loop 1 has a short timescale of $\sim$1 Myr.
It was then suggested that the N253 loop 1 is in a younger evolutionary stage than the loops 1--3.
There are young rich stellar clusters in the footpoints of the N253 loop 1 \citep[][]{gor17, ler18}.
Interestingly, the N253 loop 1 has a similar age to the cluster $\sim$1 Myr \citep[][]{ler18}, suggesting that the N253 loop 1 might be connected to the formation of the cluster.

\citet[][]{eno21} suggested that the falling gas along the field lines in the CMZ of the MW collided with the gas in the footpoint and compressed gas to formed the cluster in 0.5 Myr.
These authors argued that the molecular stream connecting the active star-forming regions in the loop connecting Sgr~A and Sgr~B2 show a loop like distribution with a 20 pc height, which may indicate magnetic flotation \citep[][]{eno21_2}. 

The N253 loop 1 shows similar spatial and kinematic signatures to the MW loops.
We therefore suggest that a couple of proto-SSCs $\#$4-7 in the CMZ of NGC~253 are possibly being driven by multiple magnetic loops at their compressed footpoints.
However, the western footpoint of the N253 loop 1 does not have proto-SSC but a maser source \citep[][]{gor17}.
The difference of the star forming activities among the eastern/western footpoints can be explained by the difference of the duration time of the gas accumulation through the loop's bridge or, in other words, the ages of the footpoints.
Considering the 1 Myr of the age of the eastern footpoint \citep[][]{ler18}, given that the age of the western foot point is younger than the eastern one, then the star forming activity is still not quite intensive right now at the western footpoint.
This possibility deserves further investigation.

In the past, the shell-like distributions and the non-rotating gas components in NGC~253 are produced by the stellar feedback  \citep[][]{sof94, sak06, bol13, wal17, zsc18, kri19, lev21, mil21}.
The present work invokes a new alternative as the origin of the gas distribution.
\citet[][]{kri19} showed that there are several more non-rotating components in the CMZ of NGC~253.
A detailed study of the features may deserve a detailed analysis similar to the present work.
Such a study (e.g., R. Konishi et al. 2022, in preparation, R. Enokiya et al. 2022, in preparation) may lead to recognition of more loop candidates, or signatures of a bar potential or IMBH, allowing us to obtain a deeper insight into the CMZs in general.\\

\section{Conclusions}\label{sec:conclusion}

We analyzed the kinematics of molecular gas in the central molecular zone of NGC~253 by using a high-resolution ALMA data set with the angular resolution of $\sim$0$\farcs$18 ($\sim$3 pc) in $\cott$, and derived molecular mass from a $\cooz$ data set with the angular resolution of $\sim$1$\farcs$6 ($\sim$30 pc).
The conclusions are summarized as follows.
\begin{enumerate}

\item We found a loop-like molecular structure exhibiting a non-circular motion toward the southwest of the galactic center of NGC~253. 
The length, width, and height of the loop-like structure are 300, 50, 20 pc, respectively.
We derived molecular mass and maximum column densities at the bases/middle region ---those are $\sim$10$^{7}$ $\sm$, $\sim$1 $\times$ 10$^{23}$ cm$^{-2}$, and $\sim$6.6 $\times$ 10$^{22}$ cm$^{-2}$, respectively--- from the CO data sets.

\item The bases of the loop-like structure show very large velocity dispersion of 50--100 $\kms$ and the middle region has the velocity gradient, which increases east to west along the loop.
The CO intensity ratios ($\ratio$) are significantly higher toward the inner part and bases of the loop-like structure.

\item It seems to be difficult to accelerate the velocity of the loop-like structure up to 200 $\kms$ by a bar potential \citep[][]{sco85} model.
A supershell model is also difficult because of its shortage of energy and the difference of the velocity structure.
A magnetic flotation model driven by Parker instability is one of a mechanism that can be responsible for the loop's huge energy and peculiar kinematics.

\item We compared the loop (N253 loop 1) with the magnetic loops in the Milky Way (loops 1--3), and found that the timescale of the N253 loop 1 is a few times shorter than the loops 1--3.
The N253 loop 1's large mass ratio between its bridge and footpoints implies that the loop is on the younger stage than that of the loops 1--3.
\end{enumerate}

We found that super star clusters having ages similar to the loop are located in the vicinity of the footpoints of the N253 loop 1.
This may be triggered by cloud-cloud collisions caused by downflows along the loop and this mechanism possibly triggered the starburst in the CMZ of the Milky Way \citep[][]{eno21}.\\

$Acknowledgements$. 
We thank (an) anonymous referee(s) for the constructive comments that improve the paper.
This research has made use of the NASA/IPAC Extragalactic Database (NED), which is operated by the Jet Propulsion Laboratory, California Institute of Technology, under contract with the National Aeronautics and Space Administration.
We thank the ALMA help desk for advising about data calibration, etc.
This paper makes use of the following ALMA data: ADS/JAO.ALMA $\#$2012.1.00108.S and $\#$2015.1.00274.S. ALMA is a partnership of ESO (representing its member states), NSF (USA), and NINS (Japan), together with NRC (Canada), MOST and ASIAA (Taiwan), and KASI (Republic of Korea), in cooperation with the Republic of Chile. The Joint ALMA Observatory is operated by ESO, AUI/NRAO and NAOJ.
We were supported by the ALMA Japan Research Grant of NAOJ ALMA Project, NAOJ-ALMA-263. This work was supported by JST SPRING, Grant Number JPMJSP2139.
\\

$Software$: CASA\citep[][]{mcm07}, astropy\citep[][]{astropy13, astropy18}, Matplotlib\citep[][]{matplotlib07}, APLpy\citep[][]{aplpy12}

\newpage
\appendix

\begin{figure}[htbp]
\centering
\includegraphics[width=140mm]{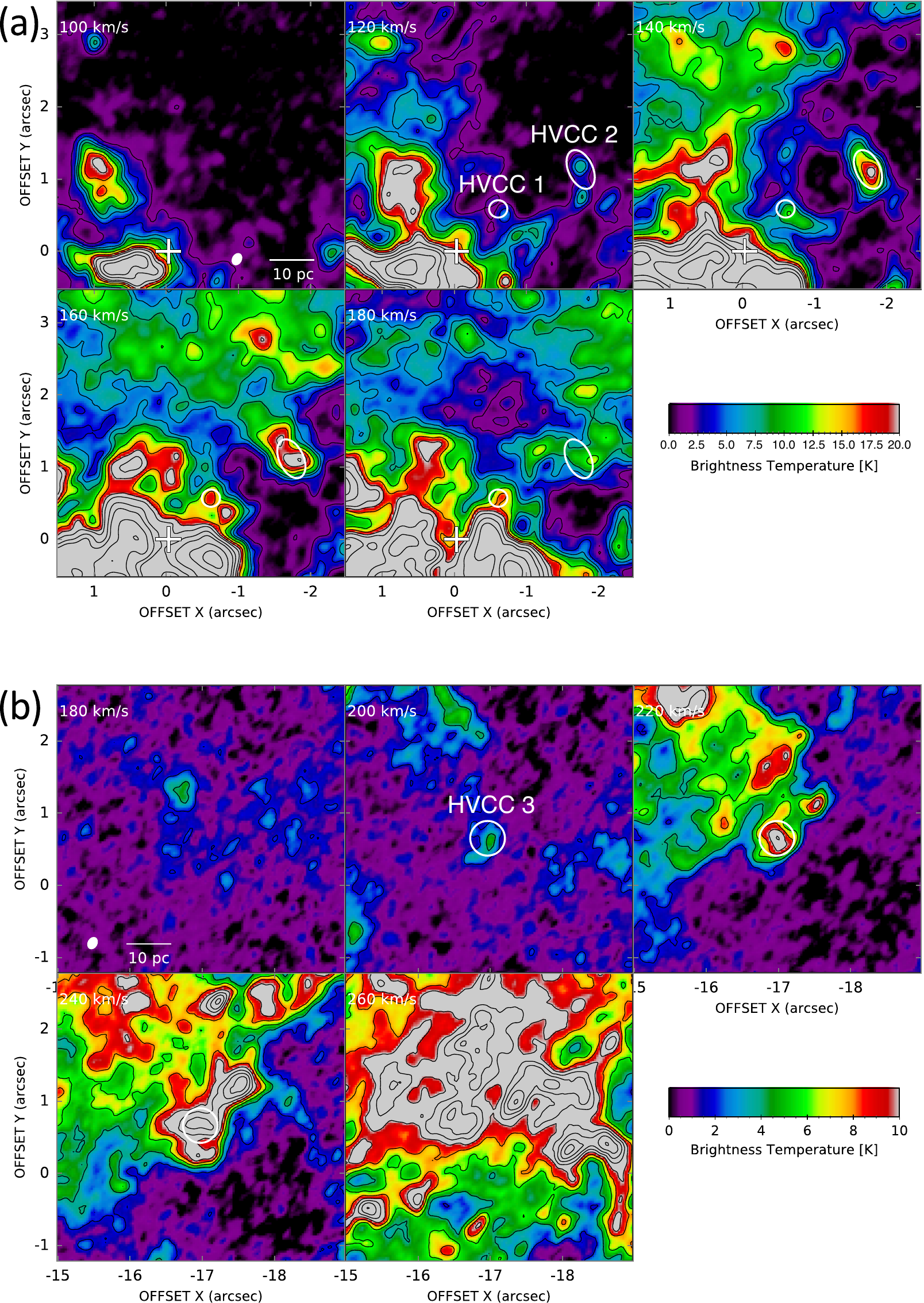}
\caption{(a) Velocity channel map toward HVCC 1 and 2 integrated every 20 $\kms$.
Two white solid ellipses at the panels at 120--180 $\kms$ indicate the positions of HVCC1 and 2.
Contours are drawn at 2, 4, 6, 8, 12, 16, 20, 24, 32, 40, 56, and 72 K.
A white cross indicates the position of the center of NGC~253.
(b) Velocity channel map toward HVCC 3 integrated every 20 $\kms$.
Two white solid ellipses on the panels at 200--240 $\kms$ indicates the position of HVCC 3.  Contours are drawn at 2, 4, 6, 8, 10, 12, 14, 16, 18, and 20 K.
A white cross is same as in (a).
\label{fig:HVCCs}}
\end{figure}

The three HVCCs mentioned in Section~\ref{result:distribution} and Figure~\ref{fig:fig_CO32_pv}b are presented in the velocity channel distributions in Figures~\ref{fig:HVCCs}a and \ref{fig:HVCCs}b.
The HVCC 1, 2, and 3 are seen in the velocity ranges of 120--186, 116--186, and 200--250 $\kms$, respectively.
We confirmed the contamination of molecular lines included in the velocity range of $\cott$ emission (i.e., HC$_{3}$N, SO, CH$_{3}$OH, NS), and no significant associated lines were found to the line emissions corresponding to the above HVCCs.

The physical parameters of the HVCCs (i.e., J2000 coordinates, positions in the OFFSET $X$ and $Y$ coordinates, velocity range, size, molecular mass, kinetic energy) are summarized in Table~\ref{tab:hvccs}.
The velocity range is measured by the p-v diagram in Figure~\ref{fig:fig_CO32_pv}b and the size is defined as the FWHM size of the moment 0 map integrated over the velocity range of each HVCC.
We then calculated the molecular mass from the moment 0 map with Eq.\ref{eq:alphafactor} and derived kinematic energy by using the velocity range.
The sizes and kinematic energies of the three HVCCs are 6--11 pc, and $\sim$10$^{51}$ erg, respectively, and thus equivalent to a typical HVCC in the MW whose size and kinematic energy are 5--10 pc, and $\sim$10$^{51}$ erg, respectively \citep[e.g.,][]{oka98}.
 The compactness and relatively higher energy of an HVCC cannot be explained by a SNe, and then the origin of them might be related to the gas dynamics of CMZs in galaxies.
 The origin of an HVCC is considered to be a multiple SNe \citep[e.g., ][]{tan07}, a Cloud-Cloud Collision \citep[e.g., ][]{sor19, eno21}, and an IMBH \citep[][]{oka16, tak19}.
The HVCCs found in the present study are limited in space and thus we still do not discuss the connection between HVCCs and gas dynamics of the CMZ of NGC~253.
Thus, further investigations of HVCCs for the entire CMZ area of NGC 253 is required and the results will be published soon (R. Enokiya et al. 2022, in preparation.)


\begin{deluxetable*}{cccc}
\tabletypesize{\scriptsize}
\tablewidth{0pt}
\tablecaption{Physical properties of the HVCCs
\label{tab:hvccs}}
\tablehead{
\colhead{}
 & 
HVCC1 & 
HVCC2 & 
HVCC3
\\[-2.5mm]
}
\startdata
R.A. & 00h47m33.081s & 00h47m32.984s & 00h47m32.090s \\
Decl. & $-$25d17m17.39s & $-$25d17m17.66s & $-$25d17m26.63s \\
OFFSET $X$ [$\arcsec$] & $-$0.60 & $-$1.83 & $-$16.92 \\
OFFSET $Y$ [$\arcsec$] & $+$0.57 & $+$1.11 & $+$0.72 \\
Velocity range [$\kms$] & 120--186 & 116--186 & 200--250 \\
FWHM velocity width [$\kms$] & 52 & 50 & 52 \\
FWHM size [pc] & 6.4 & 8.3 & 11 \\
Molecular mass [$\sm$] & 1.9 $\times$ 10$^{4}$ & 4.1 $\times$ 10$^{4}$ & 7.6 $\times$ 10$^{4}$ \\
Kinetic energy [erg] & 5.0 $\times$ 10$^{50}$ & 1.0 $\times$ 10$^{51}$ & 2.1 $\times$ 10$^{51}$ \\
\enddata
\end{deluxetable*}

\begin{figure}[htbp]	
\centering
\includegraphics[width=160mm]{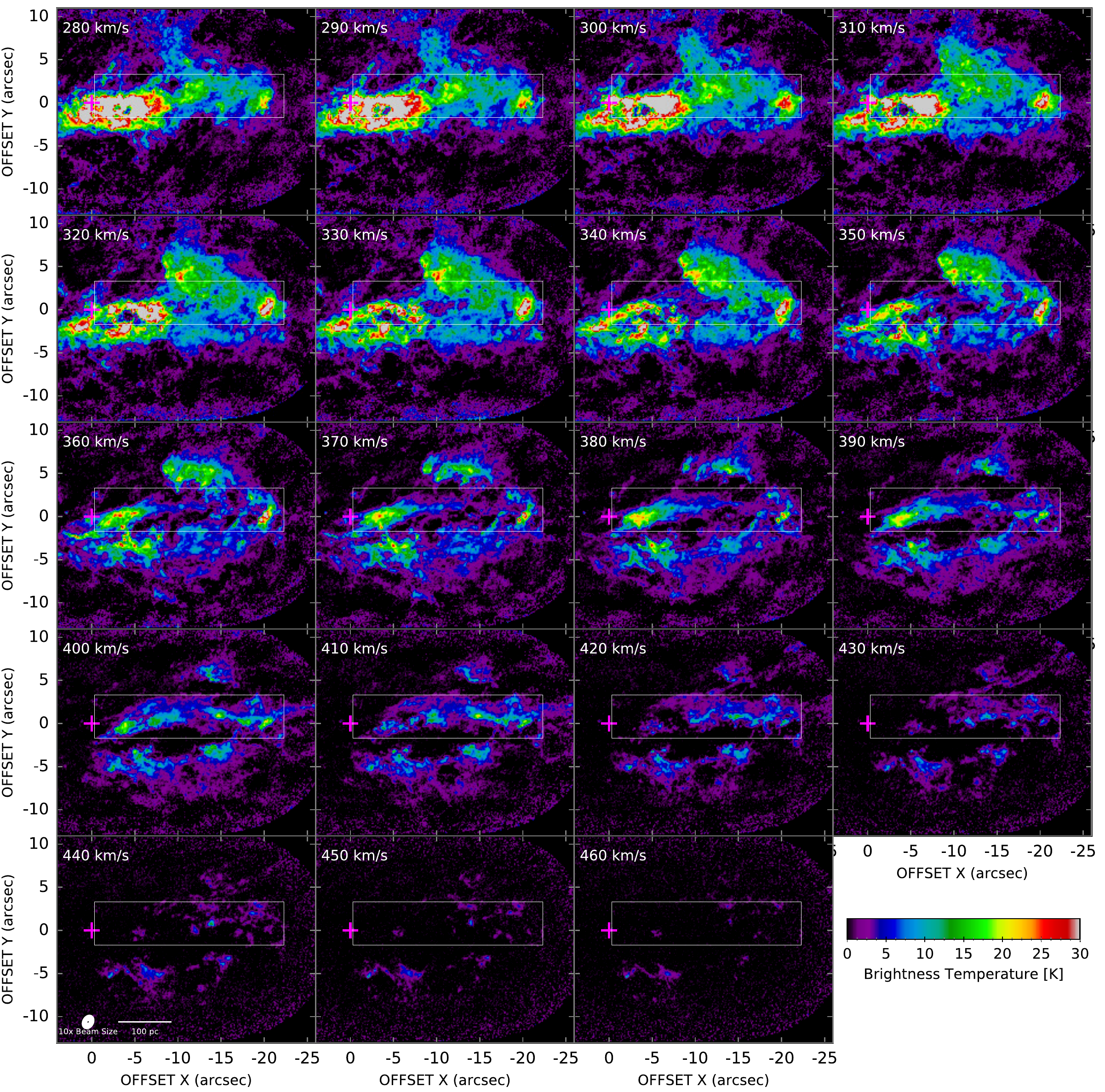}
\caption{Detailed velocity channel map of $\cott$ toward the loop structure in the OFFSET $X$ and $Y$ coordinates integrated every 10 $\kms$.
A pink cross indicates the position of the center of NGC~253.
The area including the loop is indicated by white rectangle in each panel.
\label{fig:chn_ap2}}
\end{figure}

Here we show the detailed distribution in the region of the loop-like structure (Figure~\ref{fig:chn_ap2}). Figure~\ref{fig:chn_ap2} is a channel map of $\cott$ every 10 $\kms$.
The region of the loop is shown as rectangles at every channel. 
At 280--350 $\kms$, the loop is not found but the circular motion components are dominant.
As shown in Figure~\ref{fig:fig_loop32_grid}, we found the loop at 360--440 $\kms$, which exhibits the prominent bases and the bridge with the linear velocity gradient.

\newpage
\bibliography{NGC253_loop_konishir}{}
\bibliographystyle{aasjournal}

\end{document}